\documentclass{biometrika}
\usepackage{amsmath}
\usepackage[svgnames,x11names,table]{xcolor}
\usepackage{newtxtext}
\usepackage[subscriptcorrection]{newtxmath}
\usepackage{mathrsfs}
\usepackage{booktabs}
\usepackage{threeparttable}
\usepackage{changepage}
\usepackage{multirow}
\usepackage{hyperref}
\graphicspath{{figures/}}
\makeatletter
\def\ps@plain{\let\@mkboth\@gobbletwo
  \let\@oddhead\@empty\let\@evenhead\@empty
  \def\@oddfoot{\hfil\thepage\hfil}\let\@evenfoot\@oddfoot}
\newcounter{subsubsection}[subsection]

\providecommand\subsubsection{\@startsection{subsubsection}{3}{\z@}{-9.7pt plus -2pt minus -2pt}{2pt}{\subsectionfont}}
\makeatother
\hypersetup{
  pdftitle={Distribution-Free Prediction Sets for Regression under Target Shift},
  pdfauthor={Menghan Yi, Yanlin Tang, Huixia Judy Wang},
  colorlinks=true,
  linkcolor=blue,   %
  citecolor=blue,   %
  urlcolor=blue,    %
  filecolor=blue
}
\definecolor{red}{rgb}{1,0,0}
\definecolor{green}{rgb}{0,1,0}
\definecolor{blue}{rgb}{0,0,1}

\def\boxit#1{\vbox{\hrule\hbox{\vrule\kern6pt
          \vbox{\kern6pt#1\kern6pt}\kern6pt\vrule}\hrule}}

\definecolor{judy}{rgb}{0.6, 0.2, 0.8}
\definecolor{menghan}{rgb}{0.2, 0.6, 0.8}
\definecolor{yanlin}{rgb}{0.8, 0.0, 0.0}

\def\bX{\mathbf{X}}
\providecommand{\bx}{\mathbf{x}}
\usepackage{caption}
\begin{document}
\markboth{Yi, Tang and Wang}{Distribution-Free Prediction Sets for Regression under Target Shift}
\title{Distribution-Free Prediction Sets for Regression under Target Shift}
\author{Menghan Yi\textsuperscript{1}}
\affil{\textsuperscript{1}Department of Biostatistics, University of Michigan,\\
Ann Arbor, Michigan, 48104, U.S.A.}
\author{Yanlin Tang\textsuperscript{2$\ast$}}
\affil{\textsuperscript{2}KLATASDS-MOE, School of Statistics, East China Normal University,\\
Shanghai, 200062, China}
\author{\and Huixia Judy Wang\textsuperscript{3,$\ast$}}
\affil{\textsuperscript{3}Department of Statistics, Rice University,\\
Houston, Texas, 77005, U.S.A.}
\thanks{$\ast$Corresponding authors: Yanlin Tang \href{mailto:yltang@fem.ecnu.edu.cn}{yltang@fem.ecnu.edu.cn}; Huixia Judy Wang, \href{mailto:jw322@rice.edu}{jw322@rice.edu}}
\maketitle

\begin{abstract}
In many real-world applications, statistical inference is challenged by the scarcity of labeled outcomes due to high data collection costs and technical barriers.
In the motivating MIMIC-III dataset, we aim to evaluate in-hospital illness severity among publicly insured patients, yet outcome data for this group are limited.
To address this challenge, we propose a framework for constructing efficient conformal prediction sets for target outcomes by leveraging an auxiliary source distribution with abundant labeled data that is related to the target distribution through a target shift assumption.
When target labels are unavailable, prediction relies solely on source data; when partial labels are observed, they can be incorporated to improve predictive efficiency.
To overcome data non-exchangeability and distributional non-identifiability, we identify the likelihood ratio by aligning covariate distributions of the source and target domains within a finite B-spline function space.
We further develop a weight-adjusted conditional density estimator to construct highest predictive density sets that accommodate complex error structures, including asymmetry and multimodality.
This estimator models the source conditional density via a quantile process and applies weighting transformations to approximate the target conditional density.
We establish theoretical guarantees for the proposed method and demonstrate its finite-sample performance through simulation and an application to the MIMIC-III clinical database.
\end{abstract}


\section{Introduction}
\label{sec:intro}

In many domains, such as clinical medicine and social surveys, collecting complete outcomes from the target distribution $Q$ is costly or nearly impossible, posing fundamental challenges to statistical inference. To mitigate this limitation, we leverage a related but distinct source distribution $P$ that contains abundant labeled outcome data to support the inference process. 
This strategy, however, is only viable when the relationship between the two distributions can be meaningfully characterized. 
In this paper, we assume a target shift relationship between $P$ and $Q$: the two distributions differ in the joint distribution of the outcome $Y$ and covariate $\bX$, while the conditional distribution of the covariates given the outcome, $\mathbf{X}\,|\, Y$, remains invariant across the two populations.
Target shift commonly arises in anti-causal inference settings.  For example, in medical diagnosis, symptoms are used to infer the underlying disease; although disease prevalence may vary across regions or seasons, the mechanism generating symptoms from disease is often relatively stable.

Valid inference under target shift is particularly challenging for two reasons. First, the non-exchangeability between source and target data invalidates standard prediction and inference procedures developed under identical training and test distributions. Second, the scarcity of labeled outcomes in the target domain makes it difficult to identify the shift and accurately recover the target conditional distribution. Motivated by these challenges, we study the problem of constructing prediction sets for a new individual from the target distribution $Q$. We consider settings in which labeled data are primarily available from the source distribution $P$, while the target domain provides covariates with either no labeled outcomes or only a small number of labeled outcomes, corresponding to unsupervised and semi-supervised settings, respectively.

Most existing literature on target shift primarily focuses on classification problems in machine learning, following two main lines of research: one based on expectation-maximization (EM) algorithms that maximize the likelihood of unlabeled target samples \citep{saerens2002adjusting, chan2005word, alexandari2020maximum}, and the other based on matching moments or distributions between the source and target domains to identify the shift \citep{du2014semi, lipton2018detecting, azizzadenesheli2019regularized}.
However, existing literature on regression settings with continuous outcomes, which involve infinite-dimensional parameter inference, is rather limited.
\cite{zhang2013domain} and \cite{nguyen2016continuous} addressed the infinite-dimensional challenge by projecting the density function onto a computable function subspace and performing distribution matching within this space.
\cite{lee2024doubly} proposed a flexible estimation approach using the efficient influence function, which remains valid even when the target shift or regression function is misspecified, and applies to both classification and regression settings.
However, these methods aim to estimate unknown components of the target distribution $Q$, whereas our goal is different: we directly construct prediction sets for the outcome by quantifying uncertainty, without requiring correct specification of the regression model $Y  \,|\, \mathbf{X}$.

We consider prediction set construction for the target outcome using conformal prediction (CP, \citealp{vovk2005algorithmic, lei2013distribution, lei2018distribution}), a flexible framework that forms prediction sets by retaining candidate values that conform to the observed data pattern. A key advantage of conformal prediction is that its validity does not rely on correct model specification.
However, its standard guarantees do rely on exchangeability between the calibration data and the target point. Under target shift, this requirement fails due to the distributional mismatch between $P$ and $Q$.

Several recent studies have investigated conformal prediction under distribution shift.
For example, \citet{barber2023conformal} mitigated coverage loss under non-exchangeability using data-independent weights in conformal calibration, such as pre-specified distance- or time-decaying schemes that downweight observations less similar to the target point.
\cite{gibbs2021adaptive, gibbs2024conformal} developed methods for handling arbitrary time-varying distribution shifts in an online setting, where the confidence level is adaptively updated to ensure valid prediction intervals over long time horizons.
\cite{cauchois2024robust} considered arbitrary distribution shifts by identifying a conservative quantile over an $f$-divergence ball around the training distribution to ensure valid coverage. While these approaches are applicable in principle under target shift, their accommodation of broad, unstructured distribution shifts may lead to conservative inference with wider prediction sets.

To obtain sharper predictive inference under target shift, we develop Conformal Prediction Under Target Shift (CPUTS), a new framework that explicitly addresses the non-identifiability of the target distribution and the non-exchangeability between source and target samples by leveraging target-domain covariate information. In particular, we exploit the structure of target shift to identify and estimate likelihood-ratio weights, which are then incorporated into conformal prediction to calibrate prediction sets for the target population \citep{tibshirani2019conformal, lei2021conformal, candes2023conformalized}. This is challenging because the required weights depend on aspects of the target distribution that are not directly identifiable, and are especially difficult to estimate when labeled target outcomes are scarce or unavailable.
To overcome this difficulty, we use the distribution matching implied by target shift to construct a consistent estimator of the outcome likelihood ratio, yielding valid predictive inference in both unsupervised and semi-supervised settings.
This perspective also distinguishes our setting from prior work on classification under unlabeled target shift \citep{podkopaev2021distribution, si2023pac}, as we focus on regression with continuous outcomes, where likelihood-ratio estimation becomes a function-valued problem. 
To handle this, we develop a sieve-projection approach that identifies the sieve coefficients by matching the reweighted source and target covariate distributions.
The resulting B-spline-based estimator is flexible to capture local target-shift patterns and admits new theoretical guarantees.

Beyond validity, our framework is also designed to improve efficiency. Rather than relying only on distribution-free calibration, we further construct prediction sets using the target conditional density, which can produce more compact sets than standard conformal methods \citep{vovk2005algorithmic, tibshirani2019conformal}, particularly when the conditional distribution is asymmetric, heteroscedastic, or multimodal. Because this density cannot be directly estimated without labeled target data, we propose a weighted adjustment estimator that first recovers the conditional density under $P$ from the derivative of the estimated conditional quantile function and then transfers it to $Q$ through a target-shift correction.

The remainder of the article is structured as follows. Section \ref{sec:meth} introduces the proposed method, including the problem setup, motivation, and the CPUTS procedure for both unlabeled and labeled target scenarios. 
Theoretical properties are discussed in Section \ref{sec:theory}. 
Section \ref{sec:simul} provides empirical results from simulation studies.
In Section \ref{sec:real_data}, we illustrate the application of our method using the MIMIC-III database. 
Finally, Section \ref{sec:conc} concludes the article with a discussion. 
Technical proofs and sensitivity analyses are provided in the supplementary material.

\section{Proposed Method}
\label{sec:meth}

\subsection{Setup and Motivation}\label{sec_setup}
Let $Y$ denote the univariate response variable of interest and $\mathbf{X} \in \mathbb{R}^p$ the corresponding covariate vector, jointly drawn from either the source distribution $P(\mathbf{x}, y)$ or the target distribution $Q(\mathbf{x}, y)$.
The source $P(\mathbf{x}, y)$ provides abundant labeled data, denoted as $\{(\mathbf{X}_i^P, Y_i^P)\}_{i=1}^{n_P}$, whereas the target $Q(\mathbf{x}, y)$ typically includes covariates, denoted as $\{\mathbf{X}_i^Q\}_{i=1}^{n_Q}$, with few or no corresponding labels.
Given a new individual from the target distribution with observed covariates $\mathbf{X}_{n+1}^Q$, our goal is to construct a covariate-dependent prediction set $\widehat{C}(\mathbf{X}_{n+1}^Q; \alpha)$ for the unobserved response $Y_{n+1}^Q$, so that the following coverage guarantee holds:
    \begin{equation}\label{cover-guara}
        \mathbb{P}\left\{ Y^Q_{n+1} \in \widehat{C}(\mathbf{X}^Q_{n+1}; \alpha) \right\} \geq 1 - \alpha,
    \end{equation}
where \(\alpha \in (0, 1)\) is a user-specified miscoverage level, and the probability \(\mathbb{P}\) accounts for all sources of randomness in the data.

Predicting $Y_{n+1}^Q$ is particularly challenging due to the lack of labeled data. 
To address this, we aim to leverage labeled source data for prediction, which requires an underlying relationship between the source and target distributions.
In our work, we assume that this relationship follows the target shift condition,
$$
p(y) \neq q(y) \quad \text{but} \quad p(\mathbf{x} \,|\, y) = q(\mathbf{x} \,|\, y),
$$
a setting commonly encountered in real-world applications involving inverse inference. For example, in medical diagnosis, disease prevalence may vary across regions or seasons, while the pattern of symptoms caused by diseases remains stable.
In this setting, existing work \citep{nguyen2016continuous, lee2024doubly} typically focuses on inferring specific parameters or functionals of $Q(\mathbf{x}, y)$, whereas our goal is to construct a prediction interval for $Y_{n+1}^Q$ by quantifying uncertainty.
To this end, we leverage conformal prediction (CP, \citealp{lei2013distribution, lei2018distribution}) to construct distribution-free prediction sets that quantify predictive uncertainty by assessing how well new test points conform to the observed data.
However, its coverage guarantee relies on the exchangeability assumption, which does not hold in our target shift setting.
Moreover, limited or missing labels in the target distribution pose significant challenges for identifying the underlying distribution.

In this work, we propose a method called Conformal Prediction Under Target Shift (CPUTS), which addresses the challenges of non-exchangeability and distributional non-identifiability with the aim of constructing prediction sets that satisfy
the coverage guarantee \eqref{cover-guara}.
Our method is designed to address two practical scenarios.
In \textbf{Scenario 1}, where no labels are available from target $Q(\mathbf{x}, y)$, CPUTS constructs predictions solely based on labeled data from the source $P(\mathbf{x}, y)$.
In \textbf{Scenario 2}, where a small portion of target samples is labeled, CPUTS primarily utilizes the labeled data from $Q(\mathbf{x}, y)$ to construct predictions, while incorporating $P(\mathbf{x}, y)$ data to enhance prediction efficiency.
Details of the CPUTS methods for Scenario 1 and Scenario 2 are provided in Sections \ref{unlabel_CPUTS} and \ref{label_CPUTS}, respectively.

\subsection{CPUTS on Unlabeled Data}\label{unlabel_CPUTS}
The idea of CP is to construct the prediction set by including all plausible values $y$ of $Y_{n+1}^Q$ that cannot be rejected by a level-$\alpha$ test of $H_0: Y_{n+1}^Q = y$.
To test the hypothesis, we construct a nonconformity score to evaluate the conformity of the test point $y$ with the trained model and determine whether to reject the hypothesis based on its rank.
Under distribution shift, this rank is no longer uniformly distributed, so that standard CP generally fails to attain the nominal coverage level.

\begin{figure}[htpb!]
    \centering
    \includegraphics[width=0.8\linewidth]{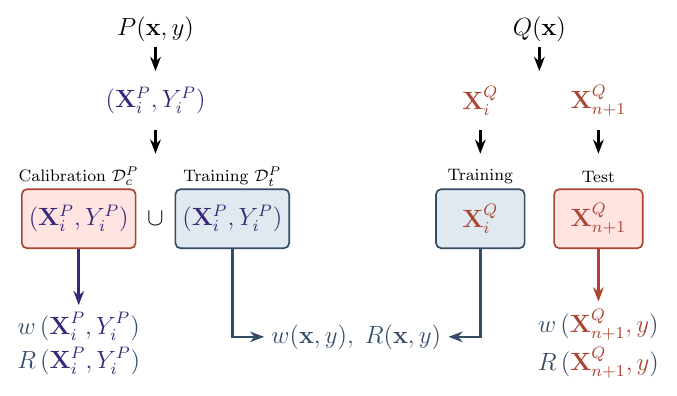}
    \caption{Data sources and sample splitting scheme for $p$-value construction in CPUTS under Scenario 1.}
    \label{fig:task1-split}
\end{figure}

Although existing methods \citep{tibshirani2019conformal, lei2021conformal, candes2023conformalized} can correct for covariate shift by reweighting samples using the density ratio between the source and target domains, these approaches do not apply directly in our setting because the target density ratio is generally not identifiable from unlabeled target data without additional assumptions.
To address this issue, we propose CPUTS, which approximates the density ratio by matching the marginal covariate distributions of the source and target domains in a sieve-based projection space.
Under the target shift assumption, this approach provides a flexible and effective way to correct for distribution mismatch and to enable accurate estimation of the nonconformity score. The resulting procedure consists of four main steps; see illustration in Figure \ref{fig:task1-split}.

\textbf{Step 1: Data Splitting.}
Recall that the available datasets are as follows:
$$
\bigl\{(\mathbf{X}_i^{P}, Y_i^{P})\bigr\}_{i=1}^{n_P}\overset{\text{i.i.d.}}{\sim}P(\mathbf{x},y), 
\qquad
\bigl\{\mathbf{X}_i^{Q}\bigr\}_{i=1}^{n_Q} \;\cup\; \mathbf{X}_{n+1}^{Q} \overset{\text{i.i.d.}} {\sim}Q(\mathbf{x}).
$$
We split the labeled $P$ data into two disjoint subsets: a training set $\mathcal{D}^P_t$ and a calibration set $\mathcal{D}^P_c$.
The set $\mathcal{D}^P_t$, together with the unlabeled $Q$ data, is used to estimate the target shift and fit the nonconformity score function, while $\mathcal{D}^P_c$ is used to compute and compare nonconformity scores for the test point.
We use a fixed training proportion, so that
$|\mathcal D_t^P|\asymp n_P$.

\textbf{Step 2: Nonconformity Score.}
Define a nonconformity score $R(\mathbf{x}, y): \mathcal{X} \times \mathcal{Y} \rightarrow \mathbb{R}$ that quantifies the degree of deviation between a test point $(\mathbf{x}, y)$ and the trained model, where a larger score indicates greater deviation.
This function can be chosen flexibly without affecting the coverage guarantee.
In CPUTS, we set
$$
R(\mathbf{x}, y; \widehat{q}) = -\widehat{q}(y \,|\, \mathbf{x}),
$$
where $\widehat{q}(y \,|\, \mathbf{x})$ denotes the estimated conditional density of $y$ given $\mathbf{x}$ under the target distribution $Q$.
This nonconformity score reflects the full structure of the target distribution and is motivated by highest predictive density sets \citep{izbicki2022cd} capable of accommodating heteroscedasticity, asymmetry, and multimodality.
Compared to standard residual-based scores \citep{shafer2008tutorial, lei2018distribution, tibshirani2019conformal}, this score can produce smaller and more informative prediction sets. 
However, conditional density estimation is particularly challenging in our setting, since no labeled target outcomes are observed. 
To address this challenge, we propose a weight-corrected estimator that first estimates the conditional density under the source distribution and then adjusts it to recover the target version through the target-shift relationship; see Section~\ref{sec: cpls_density}.

\textbf{Step 3: Evaluate the nonconformity.}
We compare the nonconformity score of the test point $y$ with those from the calibration set $\mathcal{D}^P_c$.
Since these scores are derived from different distributions and are not directly comparable, we apply the density-ratio weight \citep{tibshirani2019conformal}, defined as
\begin{equation}\label{eq:density_ratio}
w(\mathbf{x}, y) = \frac{q(\mathbf{x}, y)}{p(\mathbf{x}, y)},
\end{equation}
which corrects for non-exchangeability and yields a valid weighted ranking.
This yields a weighted $p$-value that quantifies the percentile rank of the test point’s nonconformity score:
\begin{equation}\label{p-value-task1}
p(y)
=
\sum_{i\in\mathcal D_c^P}
\varpi_i(y)
\mathbb I\{R(\mathbf X_i^P,Y_i^P)
\ge R(\mathbf X_{n+1}^Q,y)\}
+\varpi_{n+1}(y),
\end{equation}
where the weights $\varpi_i(y)$ are normalized based on the density ratio function \eqref{eq:density_ratio}:
$$
\begin{aligned}
&\varpi_i(y) =  \frac{w(\mathbf{X}^P_i, Y^P_i)}{\sum_{j\in \mathcal{D}^P_c} w(\mathbf{X}^P_j, Y^P_j)  + w(\mathbf{X}_{n+1}^Q, y) }, ~~i\in \mathcal{D}^P_{c},
\\
&\varpi_{n+1}(y) =
\frac{w(\mathbf{X}_{n+1}^Q, y)}{\sum_{j\in  \mathcal{D}^P_c} w(\mathbf{X}^P_j, Y^P_j) + w(\mathbf{X}_{n+1}^Q, y)}.
\end{aligned}
$$
If the denominator of the normalized weights is zero,
we set $p(y)=1$.
Estimating the weight function is also highly challenging due to the lack of data required to identify $q(\mathbf{x}, y)$.
To address this, we propose a distribution-matching approach that avoids explicit estimation of $q(\mathbf{x}, y)$ and instead directly recovers the ratio $w(\mathbf{x}, y)$ by aligning the covariate distributions from the source and target domains. See Section~\ref{sec: cpls_weight} for details.

\textbf{Step 4: Construct the prediction set.}
We construct the prediction set by including candidate values of $y$ with higher conformity to the trained model, as indicated by larger $p$-values: $\widehat{C}(\mathbf{X}_{n+1}^Q ; \alpha) = \{y: p(y) > \alpha\}$.

\subsubsection{Weight Estimation}\label{sec: cpls_weight}
Under the target shift assumption $q(\mathbf{x}|y) = p(\mathbf{x}|y)$, the density ratio simplifies to
$$
w(\mathbf{x}, y) = \frac{q(\mathbf{x}, y)}{p(\mathbf{x}, y)} = \frac{q(y)}{p(y)} := w_0(y),
$$
so it depends only on the univariate variable $y$, rather than on the potentially high-dimensional covariate $\bx$. This simplification reduces the estimation problem, but it remains challenging to estimate $w_0(y)$ because $q(y)$ is unobserved in the absence of labeled target data.
To overcome this challenge, we adopt a sieve-basis projection combined with the idea of distribution matching \citep{zhang2013domain, nguyen2016continuous, guo2020ltf}, which directly estimates $w_0(y)$ by aligning the observed covariate distribution in the target domain with that implied by the source domain.

Define the reweighted distribution $q_{w}(\mathbf{x}) := \int w(y) p(\mathbf{x}, y) dy$ with respect to the source distribution.
Under the target shift assumption, if the weight function takes its true form $w_0(y)=q(y) / p(y)$, then $q_{w_0}(\mathbf{x}) = q(\mathbf{x})$. 
Therefore, $w_0(y)$ can be estimated by minimizing the discrepancy between $q_w(\mathbf{x})$ and $q(\mathbf{x})$, subject to the non-negativity constraint $w(y) \geq 0$ and the normalization constraint $\int w(y) p(y) d y=1$.
However, this minimization problem is infinite-dimensional and requires estimating the unknown densities $q(\mathbf{x})$ and $p(\mathbf{x}, y)$, which may be high-dimensional.
To address these challenges, we approximate both the weight function $w(y)$ and the density difference $q_w(\mathbf{x})-q(\mathbf{x})$ using basis expansions: 
\begin{equation}\label{eq:bases}
w(y) \approx \sum_{j=1}^{J_n} \alpha_j B_{j}(y) = \boldsymbol{\alpha}^{\top} \mathbf{B}(y), \quad 
q_w(\mathbf{x})-q(\mathbf{x}) \approx \sum_{k=1}^{K_n} \beta_k \psi_k(\mathbf{x}) = \boldsymbol{\beta}^{\top} \boldsymbol{\psi}(\mathbf{x}),
\end{equation}
where $\mathbf{B}(y) = [B_{1}(y), \ldots, B_{J_n}(y)]^{\top}$ and
$\boldsymbol{\psi}(\mathbf{x}) = [\psi_1(\mathbf{x}), \ldots, \psi_{K_n}(\mathbf{x})]^{\top}$
are vectors of $J_n$ and $K_n$ basis functions, respectively, and $\boldsymbol{\alpha} \in \mathbb{R}^{J_n}$, $\boldsymbol{\beta} \in \mathbb{R}^{K_n}$ are the corresponding coefficient vectors.
For the response variable, we employ univariate B-spline basis functions of degree $d$ \citep{schumaker2007spline}, denoted by $B_j(y) := B_{j,d}(y)$, to flexibly accommodate diverse target-shift patterns.
For the covariate space, we adopt Gaussian radial basis functions \citep{buhmann2000radial},
defined as
$\psi_k(\mathbf{x}) = \exp\!\left\{-\|\mathbf{x}-\boldsymbol{\mu}_k\|^2/(2\sigma_k^2)\right\}$,
where $\boldsymbol{\mu}_k$ and $\sigma_k$ denote the center and bandwidth, respectively.
This choice leads to closed-form expressions for a key inner-product matrix and thereby simplifies computation in high-dimensional settings.

To estimate $\boldsymbol{\beta}$ in \eqref{eq:bases}, we minimize the integrated squared error between the true difference function $q_w(\mathbf{x})-q(\mathbf{x})$ and its approximation via the basis expansion:
$$
\int \left[ \boldsymbol{\beta}^{\top} \boldsymbol{\psi}(\mathbf{x}) - \left\{q_w(\mathbf{x})  - q(\mathbf{x}) \right\} \right]^2 \mathrm{d}\mathbf{x} 
=  \boldsymbol{\beta}^\top \boldsymbol{U} \boldsymbol{\beta} - 2\boldsymbol{\beta}^\top \left( \boldsymbol{V} \boldsymbol{\alpha} - \boldsymbol{u} \right) + C,
$$
where   
$\boldsymbol{U} = \int \boldsymbol{\psi}(\mathbf{x})\boldsymbol{\psi}^\top(\mathbf{x})\, \mathrm{d}\mathbf{x}$,
$\boldsymbol{V} = \mathbb{E}_{(\mathbf{X}, Y) \sim p(\mathbf{x}, y)}[\boldsymbol{\psi}(\mathbf{X}) \mathbf{B}^\top(Y)]$,
 $\boldsymbol{u} = \mathbb{E}_{\mathbf{X} \sim q(\mathbf{x})}[\boldsymbol{\psi}(\mathbf{X})]$,
and the term $C$ is a constant independent of $\boldsymbol{\beta}$.
The equality holds because $q_w(\mathbf{x})$ is approximated by $\boldsymbol{\alpha}^{\top} \int \mathbf{B}(y)\, p(\mathbf{x}, y)\, \mathrm{d}y$.
With a ridge penalty, we can profile out $\boldsymbol{\beta}$
and obtain the following closed-form solution:
$$
\widehat{\boldsymbol{\beta}}(\boldsymbol{\alpha})
=(\boldsymbol{U}+\delta_n \boldsymbol{I})^{-1}(\widehat{\boldsymbol{V}}\boldsymbol{\alpha} - \widehat{\boldsymbol{u}}),
$$
where $\widehat{\boldsymbol{V}} = \frac{1}{|\mathcal{D}^P_t|} \sum_{i\in \mathcal{D}^P_t} \boldsymbol{\psi}(\mathbf{X}_i^P) \mathbf{B}^\top(Y_i^P)$,
$\widehat{\boldsymbol{u}} = \frac{1}{n_Q} \sum_{i=1}^{n_Q} \boldsymbol{\psi}(\mathbf{X}_i^Q)$,
and $\delta_n$ is a regularization parameter introduced to enhance numerical stability and prevent overfitting.
Finally, to estimate $\boldsymbol{\alpha}$, we consider
the integrated squared error
$\int \{q_w(\mathbf{x})-q(\mathbf{x})\}^2\,\mathrm{d}\mathbf{x}$,
which is approximated by
$\widehat{\boldsymbol{\beta}}(\boldsymbol{\alpha})^\top
\boldsymbol{U}
\widehat{\boldsymbol{\beta}}(\boldsymbol{\alpha})$.
We adopt a regularized quadratic criterion,
leading to the following optimization problem:
\begin{equation}\label{eq:estimator_w}
\widehat{\boldsymbol{\alpha}} = \underset{\boldsymbol{\alpha}}{\operatorname{argmin}} 
\bigg[
\boldsymbol{\alpha}^{\top}\widehat{\boldsymbol{V}}^{\top} (\boldsymbol{U}+\delta_n \boldsymbol{I})^{-1} \widehat{\boldsymbol{V}}\boldsymbol{\alpha}
-
2\widehat{\boldsymbol{u}}^{\top} (\boldsymbol{U}+\delta_n \boldsymbol{I})^{-1} \widehat{\boldsymbol{V}}\boldsymbol{\alpha}
+ {\rho_n} \boldsymbol{\alpha}^{\top} \boldsymbol{\alpha}
\bigg],
\end{equation}
subject to the constraints $\boldsymbol{\alpha}^{\top} \mathbf{B}(y) \geq 0$
and
$\boldsymbol{\alpha}^{\top}\widehat{\mathbf m}=1$,
where
$\widehat{\mathbf m}
=|\mathcal D_t^P|^{-1}
\sum_{i\in\mathcal D_t^P}\mathbf B(Y_i^P)$
and $\rho_n$ is a regularization parameter.
Hence, by substituting \eqref{eq:bases}, we obtain the estimated weight function $\widehat{w}(y) = \widehat{\boldsymbol{\alpha}}^\top \mathbf{B}(y)$.

In our approach, we employ B-spline basis functions to flexibly model the complex shape of $w(y)$ arising from various types of distribution shift. Compared to global basis functions \citep{zhang2013domain, nguyen2016continuous}, the compact support of B-splines allows for more precise local control, offering a distinct advantage in modeling localized variation in practice.
Moreover, the use of Gaussian radial basis functions facilitates computation over the covariate space, as the $L^2$ inner product matrix $\boldsymbol{U} = \int \boldsymbol{\psi}(\mathbf{x}) \boldsymbol{\psi}^\top(\mathbf{x}) \mathrm{d}\mathbf{x}$ admits a closed-form expression, eliminating the need for numerical integration.

\subsubsection{Conditional Density Estimation}\label{sec: cpls_density}
To estimate the target conditional density, a natural and intuitive strategy is to first estimate the source density using the labeled data and then adjust it based on the relationship between the two distributions.
Specifically, we construct the following reweighted density estimator via Bayes’ rule:
\begin{equation}\label{eq:dens_p}
\widehat{q}(y\,|\, \mathbf{x})  
	:= \frac{\widehat{p}(y\,|\,\mathbf{x})\widehat{w}(y)}{\int \widehat{p}(y\,|\,\mathbf{x})\widehat{w}(y)dy},
\end{equation}
where $\widehat{w}(y)$ denotes the estimated density ratio $q(y) / p(y)$ obtained in Section~\ref{sec: cpls_weight}, and $\widehat{p}(y \,|\, \mathbf{x})$ is the conditional density under source distribution $P$.

Standard methods for estimating the conditional density
$p(y\mid\mathbf x)$, including kernel and $k$-nearest-neighbor
methods, often suffer from the curse of dimensionality
\citep{hastie2009elements}.
To alleviate this problem, we adopt a semi-parametric approach
that reconstructs the conditional density through a quantile
process based on a working model,
avoiding direct smoothing over the
high-dimensional covariate space.
Importantly, the validity of the proposed CPUTS procedure
does not rely on the correct specification of this model.
Let $\xi_P(\tau \,|\, \mathbf{x})$ denote the $\tau$-th conditional quantile of $Y$ given $\mathbf{x}$ under distribution $P$.
For a continuous distribution, the conditional density $p(y \,|\, \mathbf{x})$, evaluated at the point $y  = \xi_P(\tau |  \mathbf{x})$, satisfies
$p\big\{\xi_P(\tau|\mathbf{x}) \,|\, \mathbf{x}\big\}=\{{\mathrm{d} \xi_P(\tau \,|\, \mathbf{x})}/{\mathrm{d} \tau}\}^{-1}.$
Therefore, we construct the quotient estimator \citep{siddiqui1960distribution} as
\begin{equation}\label{eq:quotient}
\widehat{p}\big\{\widehat{\xi}_P(\tau| \mathbf{x}) \,|\, \mathbf{x}\big\}=\frac{2 h_n}{\widehat{\xi}_P(\tau+h_n \,|\, \mathbf{x})-\widehat{\xi}_P(\tau-h_n \,|\, \mathbf{x})},
\end{equation}
where $h_n$ is a bandwidth that shrinks to zero as $n \rightarrow \infty$ and $\widehat{\xi}_P(\tau \,|\, \mathbf{x})$ denotes the estimated conditional $\tau$-quantile under the source distribution $P$.
The quantile estimates can be obtained using any appropriate quantile regression method under a working model, such as linear quantile regression \citep{koenker2005quantile} or quantile random forests \citep{athey2019generalized}.
To estimate the full conditional density $\widehat{p}(y \,|\, \mathbf{x})$,  we first specify a set of quantile levels  $\{\tau_1^P, \ldots, \tau_{\kappa^P}^P\}$ and compute density estimates at the corresponding estimated quantile values $\{\widehat{\xi}_P(\tau_1^P \,|\, \mathbf{x}), \ldots, \widehat{\xi}_P(\tau_{\kappa^P}^P \,|\, \mathbf{x})\}$ using equation~\eqref{eq:quotient}. 
These pointwise estimates are then interpolated, for example, linearly or using splines, to approximate $\widehat{p}(y \,|\, \mathbf{x})$ for any value of $y$.

\subsection{CPUTS on Labeled Data}\label{label_CPUTS}
\begin{figure}[htpb!]
    \centering
    \includegraphics[width=0.8\linewidth]{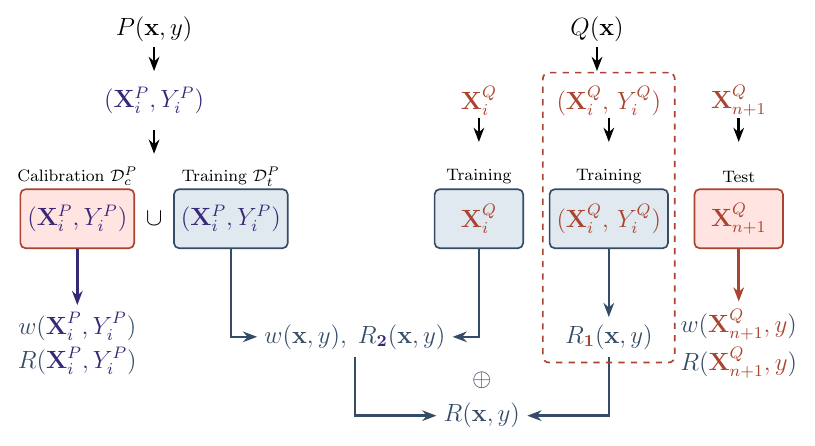}
        \caption{Data sources and sample splitting scheme for $p$-value construction in CPUTS under Scenario 2. The dashed box denotes all labeled target data.}
    \label{fig:task2-split}
\end{figure}

Suppose we have labeled samples from the source distribution and partially labeled samples from the target distribution, where only the first $n_Q^{\mathrm{lab}}$ out of $n_Q$ target samples have labels:
$$
\bigl\{(\mathbf{X}_i^{P}, Y_i^{P})\bigr\}_{i=1}^{n_P}\overset{\text{i.i.d.}}{\sim}P(\mathbf{x},y), 
\qquad
\bigl\{(\mathbf{X}_i^{Q}, Y_i^{Q})\bigr\}_{i=1}^{n_Q^{\mathrm{lab}}}\;\cup\;
\bigl\{\mathbf{X}_j^{Q}\bigr\}_{j=n_Q^{\mathrm{lab}}+1}^{n_Q}
\;\cup\; \mathbf{X}_{n+1}^{Q}
\overset{\text{i.i.d.}}{\sim}Q(\mathbf{x},y).
$$
Under this setting, a natural strategy is to conduct statistical inference primarily based on the $Q$ data, while incorporating the $P$ data as auxiliary information to enhance predictive efficiency.
In principle, different data-splitting schemes can be adopted for the $P$ and $Q$ samples to allocate them to model training and calibration. 
However, predictive efficiency of conformal prediction depends primarily on accurate estimation of nonconformity scores \citep{dunn2023distribution, hu2024two};
therefore, we allocate all labeled $Q$ data to the training set so the calibration set includes only auxiliary $P$ data. %
 The detailed splitting strategy is shown in Figure~\ref{fig:task2-split}. Our preliminary studies suggest that this strategy yields greater efficiency gains than alternative data-splitting approaches.

Specifically, we adopt a combined nonconformity score defined as
$$
R(\mathbf{x}, y) = (1-\gamma)R_1(\mathbf{x}, y) + \gamma R_2(\mathbf{x}, y),
$$
where both $R_1$ and $R_2$ take the form $-\widehat{q}(y\mid \mathbf{x})$, 
but are fitted using different training data: $R_1$ uses $\widehat{q}$ trained on 
labeled data from $Q$, whereas $R_2$ uses $\widehat{q}$ defined in \eqref{eq:dens_p} 
as in Scenario~1, fitted on labeled data from $P$ and reweighted to recover the 
target distribution $Q$.
The parameter $\gamma \in [0,1]$ controls the relative contributions of the two scores.
Our CPUTS procedure then follows Scenario 1 by using the same $p$-value \eqref{p-value-task1} to construct the prediction region, while retaining the density-ratio weight \eqref{eq:density_ratio} unchanged.
This is because the entire calibration set is drawn from $P$ and therefore must be reweighted to account for the distributional shift to $Q$.
Importantly, since the weight function in the $p$-value depends only on the change of measure from $P$ to $Q$ and not on the nonconformity score, replacing the score in Scenario 2 does not affect the weights.

To motivate the choice of $\gamma$, consider a simple case where
$R_1$ and $R_2$ are independent for a fixed $(\mathbf x,y)$,
with variances $\sigma_1^2$ and $\sigma_2^2$ over the training data.
The variance of the combined score is then
$\operatorname{Var}(R)
=(1-\gamma)^2\sigma_1^2+\gamma^2\sigma_2^2$,
which is minimized at
$\gamma^*=\sigma_1^2/(\sigma_1^2+\sigma_2^2)$.
If the two variances are inversely proportional to their
training sample sizes with the same proportionality constant,
then $\gamma^*$ becomes
$|\mathcal D_t^P|/(n_Q^{\mathrm{lab}}+|\mathcal D_t^P|)$.
This suggests a possible heuristic choice of
$\gamma=|\mathcal D_t^P|/(n_Q^{\mathrm{lab}}+|\mathcal D_t^P|)$.
Alternatively, $\gamma$ may be specified as another value
in $[0,1]$ based on practical considerations.

To construct $R_1(\mathbf{x}, y)=-\widehat{q}(y\,|\, \mathbf{x})$ based on the  $Q$ data, we directly estimate $\widehat{q}(y \,|\, \mathbf{x})$ using the quotient estimator based on its labeled samples.
Similar to \eqref{eq:quotient}, the conditional density at $\xi_Q(\tau \,|\, \mathbf{x})$, which is the $\tau$-th conditional quantile of $Y$ given $\mathbf{x}$ under distribution $Q$, can be estimated as:
$$
\widehat{q}\{\widehat{\xi}_Q(\tau \,|\, \mathbf{x}) \,|\, \mathbf{x}\} 
= \frac{2 h_n}{\widehat{\xi}_Q(\tau + h_n \,|\, \mathbf{x}) - \widehat{\xi}_Q(\tau - h_n \,|\, \mathbf{x})},
$$
where $h_n$ is a bandwidth tending to zero as $n \rightarrow \infty$, and $\widehat{\xi}_Q(\tau \,|\, \mathbf{x})$ is the estimated $\tau$-th conditional quantile.
Therefore, $\widehat{q}(y \,|\, \mathbf{x})$ can be obtained from the estimated quantile process 
$\{\widehat{\xi}_Q(\tau_1^Q \,|\, \mathbf{x}), \ldots, \widehat{\xi}_Q(\tau_{\kappa^Q}^Q \,|\, \mathbf{x})\}$,
where $\{\tau_1^Q, \ldots, \tau_{\kappa^Q}^Q\}$ is a pre-specified set of quantile levels.

\section{Theoretical Properties}\label{sec:theory}
In this section, we first establish that the proposed CPUTS method guarantees coverage under a single condition, where the coverage accuracy depends on the precision of the weight estimation.
The asymptotic properties of the weight estimator are presented later under standard regularity conditions.

\begin{assumption}\label{ass:finite_wt}
The target outcome distribution $Q_Y$ is absolutely continuous
with respect to the source outcome distribution $P_Y$,
i.e., $Q_Y\ll P_Y$.
\end{assumption}

Assumption~\ref{ass:finite_wt} implies that the support of
the target outcome distribution is contained in that of
the source outcome distribution.
This may be reasonable when the source population covers a
broader range of outcomes.
Under target shift, it also implies $Q\ll P$
for the joint distributions of $(\mathbf X,Y)$.

Let $\mathcal D_t$ denote all training data used to construct
$R$ and $\widehat w$. We assume that the calibration sample
and the target test point are independent of $\mathcal D_t$
and of each other.
\begin{theorem}\label{th:cover}
Under Assumption~\ref{ass:finite_wt}, suppose that
$\widehat w(y)\ge 0$ for $P_Y$-almost every $y$ and
$0<\mathbb E_{Y\sim P_Y}\widehat w(Y)<\infty$
hold almost surely.
Then, for any $\alpha\in(0,1)$, the CPUTS method satisfies
\begin{equation}\label{eq:cover}\mathbb P\{Y_{n+1}\in\widehat C(\mathbf X_{n+1};\alpha)
\mid\mathcal D_t\}
\ge 1-\alpha
-\frac12\mathbb E\Big[
\big|
\frac{\widehat w(Y)}
{\mathbb E_{Y'\sim P_Y}\widehat w(Y')}
-w_0(Y)
\big|
\mid\mathcal D_t
\Big],
\quad\text{a.s.},
\end{equation}
where $Y\sim P_Y$ is independent of $\mathcal D_t$.
\end{theorem}

The proof of Theorem \ref{th:cover} is provided in the supplementary material.
Theorem \ref{th:cover} demonstrates that when the target shift mechanism is known, the CPUTS method guarantees an exact coverage lower bound of $1-\alpha$.
Alternatively, when the mechanism is unknown, the coverage bound includes an additional error term depending on the accuracy of weight estimation.
We will show in Theorem~\ref{thm:weight_rate} that under standard regularity conditions and appropriate tuning parameter choices, the estimation error of the proposed estimator is asymptotically negligible.
The CPUTS method satisfies the coverage lower bound in \eqref{eq:cover} for both the unsupervised (Section~\ref{unlabel_CPUTS}) and semi-supervised (Section~\ref{label_CPUTS}) target settings.
In the semi-supervised setting, additional labeled target data can shorten prediction intervals while maintaining coverage through our data-splitting strategy, which allocates the labeled target data entirely to estimating the nonconformity scores.

Before presenting the asymptotic properties of the weight estimator, we introduce some regularity conditions.

\begin{assumption}[Weight Function Approximation]\label{assump:spline}
Let $I=[a,b]$ be a fixed bounded interval such that
$Q_Y(I)=1$.
The weight function $w_0(y)$ belongs to the Sobolev space
$W^{r,2}([a,b])$ with $r>1/2$.
We approximate $w_0(y)$ using standard normalized B-spline
basis functions $\{B_{j,d}(y)\}_{j=1}^{J_n}$ of fixed integer
degree $d$, satisfying $B_{j,d}(y)\ge0$ and
$\sum_{j=1}^{J_n}B_{j,d}(y)=1$ for all $y\in[a,b]$.
Each B-spline basis function is extended by zero outside $I$.
Define the sieve space
$\mathcal W_{J_n}
=\{w_\alpha(y)=\alpha^\top\mathbf B(y):
\alpha\in\mathbb R^{J_n}\}$.
Assume that (a) $d\ge\lceil r\rceil$; and
(b) the nonzero knot spans are quasi-uniform,
with a mesh-ratio bound independent of $n$.
\end{assumption}

\begin{assumption}[Operator Smoothing]\label{assump:Gaussian}
Define the conditional expectation operator
$(\mathcal R z)(y)
=\mathbb E_P[z(\mathbf X)\mid Y=y]
=p_Y(y)^{-1}\int_{\mathbb R^p}p(\mathbf x,y)z(\mathbf x)\,d\mathbf x$.
There exists a constant $C_{\mathcal R}>0$,
independent of $n$, such that
$\|\mathcal R z\|_{W^{\nu,2}([a,b])}
\le C_{\mathcal R}\|z\|_{L^2(\mathbb R^p)}$
for all $z\in L^2(\mathbb R^p)$,
where $0<\nu\le d+1$ is the exponent
in Assumption~\ref{assump:ident}.
\end{assumption}

\begin{assumption}\label{assump:ident}
Let $\delta_n,\rho_n\ge 0$ and
$\delta_n,\rho_n\to 0$ in \eqref{eq:estimator_w}.
Assume that $\mathbf U$ is positive definite.
Let $\mathbf G_J=\int_a^b\mathbf B(y)\mathbf B^\top(y)\,dy$.
There exist constants $c_\kappa>0$ and $\nu>1/2$,
independent of $n$, such that 
$\sigma_{\min}(
(\mathbf U+\delta_n\mathbf I_{K_n})^{-1/2}
\mathbf V\mathbf G_J^{-1/2})
\ge c_\kappa J_n^{-\nu}$,
where $\sigma_{\min}(\mathbf A)
=\inf_{\|\mathbf v\|_2=1}\|\mathbf A\mathbf v\|_2$.
\end{assumption}

\begin{assumption}[Asymptotic growth rates]\label{assump:dim_min}
Let $n=\min(n_P,n_Q)\to\infty$.
Assume that $J_n\to\infty$,
$J_n^{2\nu+1}K_n/n\to0$, and
$\rho_nJ_n^{2\nu+1}\to0$,
where $\nu$ is the exponent in
Assumption~\ref{assump:ident}.
\end{assumption}

\begin{assumption}[Density boundedness]\label{assump:py_bd}
There exist constants $0<\underline{c}_Y\le \overline{c}_Y<\infty$ such that
$\underline{c}_Y \le p_Y(y)\le \overline{c}_Y$ for all $y\in[a,b]$.
In addition, there exists a constant $0<c_{XY}<\infty$ such that
$p(\mathbf x,y)\le c_{XY}$
for almost every $(\mathbf x,y)\in\mathbb R^p\times[a,b]$.
\end{assumption}

Under Assumption~\ref{assump:spline}, we can approximate $w_0$
using B-splines with nonnegative coefficients and
an error of order $J_n^{-r}$ for sufficiently large $J_n$,
provided that $w_0$ is bounded away from zero.
Assumption~\ref{assump:Gaussian} helps control the error
in matching the target covariate distribution with the
reweighted source distribution when $w_0$ is approximated
using B-splines.
Instead of requiring the Gaussian RBFs to approximate
the entire density discrepancy,
Assumption~\ref{assump:ident} requires the moment equations
linking the source and target distributions to contain
enough information to distinguish different weight functions
within the B-spline space.
The exponent $\nu$ controls how quickly the smallest singular
value is allowed to decrease as $J_n$ increases.
A smaller $\nu$ requires the smallest singular value to
decrease more slowly, making it easier to recover
the weight function from these moments and improving
the convergence rate in Theorem~\ref{thm:weight_rate}.
Assumption~\ref{assump:ident} can be satisfied in typical settings.
For example, Proposition~2 in the Supplementary Material
shows that it holds under a simple Laplace model.
Assumption~\ref{assump:dim_min} allows the number of basis
functions to grow with the sample size while controlling
the errors in estimating the source and target moments.
It also ensures that the regularization error vanishes
as the sample size increases.
Assumption~\ref{assump:py_bd} allows us to control
the weight estimation error under both the source
distribution and the usual $L^2$ norm, while the
density bounds help control the conditional
expectations used in the proof.

\begin{theorem}\label{thm:weight_rate}
Suppose Assumptions~\ref{assump:spline},
\ref{assump:Gaussian}, \ref{assump:ident},
\ref{assump:dim_min}, and \ref{assump:py_bd} hold,
and $w_0(y)\ge c_0>0$ for all $y\in[a,b]$.
Assume that the source and target samples are
independent and i.i.d. from $p(\mathbf x,y)$
and $q(\mathbf x)$, respectively, and that
the basis functions and regularization parameters
are deterministic.
Then the estimator $\widehat w$ in
\eqref{eq:estimator_w}, with empirical normalization
$\boldsymbol\alpha^\top\widehat{\mathbf m}=1$,
satisfies
\[
\|\widehat w-w_0\|_{L^2([a,b])}^2
=
O_p\left(
J_n^{-2r}
+\frac{J_n^{2\nu}K_n}{n}
+\rho_n^2J_n^{4\nu+2}
\right),
\]
where $n=\min(n_P,n_Q)$.
\end{theorem}

The proof of Theorem~\ref{thm:weight_rate} is provided in the
supplementary material.
The upper bound consists of three terms:
(i) the error from approximating $w_0$ using B-splines,
$J_n^{-2r}$;
(ii) the error from estimating the source and target moments
using finite samples, $J_n^{2\nu}K_n/n$;
and (iii) the error due to regularization,
$\rho_n^2J_n^{4\nu+2}$.
The second term also depends on $\nu$, which controls
how quickly the smallest singular value may decrease
as $J_n$ increases.
The bound does not include separate terms for Gaussian RBF
approximation or $\delta_n$, because Assumption~\ref{assump:ident}
controls how well the finite Gaussian RBF moments identify
the weight function and accounts for $\delta_n$ through
the singular-value condition.
Under Assumption~\ref{assump:dim_min}, all three terms
converge to zero, ensuring consistency of $\widehat w$.
For example, suppose the assumptions of
Theorem~\ref{thm:weight_rate} hold with
$K_n\asymp J_n$ and
$\rho_n=o(J_n^{-(r+2\nu+1)})$.
Balancing the first two terms in the upper bound gives
$J_n\asymp n^{1/(2r+2\nu+1)}$, yielding
\[
\|\widehat w-w_0\|_{L^2([a,b])}^2
=
O_p\big(n^{-2r/(2r+2\nu+1)}\big).
\]
In Section~S2 of the supplementary material,
we examine how $J_n$, $K_n$, $\delta_n$,
and $\rho_n$ affect predictive performance
and provide practical recommendations for their selection.

\section{Simulation Study}\label{sec:simul}
In this section, we evaluate the finite-sample performance of CPUTS under both unsupervised and semi-supervised target settings. Specifically, we consider two types of distribution shift under linear and nonlinear models, and assess performance in terms of coverage probability and average interval length.

To capture different forms of target shift, we consider two settings: (i) a \textbf{location shift}, where $Y^P \sim N(0,1)$ and $Y^Q \sim N(1,1)$; and (ii) a \textbf{location–scale shift}, where $Y^P \sim N(0,1.5^2)$ and $Y^Q \sim N(1,0.5^2)$. In both settings, the conditional distribution of $X$ given $Y$ remains the same across the source and target populations. Specifically, we consider two data-generating models: (i) a \textbf{linear model}, $X=Y+\varepsilon$, and (ii) a \textbf{nonlinear model}, $X=U+\varepsilon$, where $U\mid Y\sim\operatorname{Lognormal}(Y,1)$. In both models, $\varepsilon$ is independent of $Y$ and follows a standardized skew-normal distribution with shape parameter 15.
We generate \(n_P\) labeled source samples \((X^P,Y^P)\) and \(n_Q=n_Q^{\mathrm{unlab}}+n_Q^{\mathrm{lab}}\) target samples, where \(n_Q^{\mathrm{unlab}}\) and \(n_Q^{\mathrm{lab}}\) denote the numbers of unlabeled and labeled target samples, respectively.
In these experiments, we fix $n_Q^{\mathrm{unlab}}=1000$. For the unsupervised setting, we set $n_Q^{\mathrm{lab}}=0$ and vary $n_P$ over $\{500,1000,2000,3000,4000\}$, whereas for the semi-supervised setting, we fix $n_P=2000$ and vary $n_Q^{\mathrm{lab}}$ over $\{20,50,100,300,500\}$. Each configuration is evaluated over 1000 replications, with 300 target test points per replication.

For weight estimation, we use B-spline basis functions
with equally spaced interior knots for the response
and Gaussian radial basis functions for the covariates.
For the location-shift setting, we use degree-one
B-splines with $J_n=5$.
For the location--scale shift, we use degree-two
B-splines to allow greater flexibility,
with the number of interior knots chosen based on
$c n_{P,\mathrm{tr}}^{1/5}$,
where $c\in[4,6]$ and $n_{P,\mathrm{tr}}$ denotes
the number of source training samples. 
For the Gaussian radial basis functions,
we generally set $K_n=5$, but increase it to
$K_n=14$ in the unsupervised nonlinear location-shift
setting to provide a richer basis for matching
the source and target covariate distributions.
The RBF centers are selected based on the source
training and unlabeled target covariates.
The baseline bandwidth is determined using
Silverman's rule of thumb,
\[
\sigma = 1.06 \min\bigg\{
\widehat{\sigma}_X,
\frac{\widehat{\mathrm{IQR}}_X}{1.349}
\bigg\} n_{P,\mathrm{tr}}^{-1/5},
\]
where $\widehat{\sigma}_X$ and
$\widehat{\mathrm{IQR}}_X$ denote the sample
standard deviation and interquartile range
of the source training covariates, respectively. 
To stabilize weight estimation, we use ridge penalties
$\delta_n$ and $\rho_n$ ranging from
$10^{-6}$ to $10^{-2}$.
The B-spline coefficients are constrained to be
nonnegative, with the estimated weights empirically
normalized over the source training samples.
For conditional density estimation, we employ linear quantile regression to estimate the conditional quantile process. The quantile levels are defined as $\tau_j^P = j/(\kappa^P+1)$ for $j = 1,\ldots,\kappa^P$ and $\tau_j^Q = j/(\kappa^Q+1)$ for $j = 1,\ldots,\kappa^Q$, where $\kappa^P$ and $\kappa^Q$ are chosen based on $2n_{P,\mathrm{tr}}^{1/3}$ and $2(n_Q^{\mathrm{lab}})^{1/3}$, respectively, with both constrained to lie in $[9,25]$. The bandwidth $h_n$ is selected using the Hall--Sheather rule implemented in the \textit{quantreg} package.

In the unsupervised setting with $n_Q^{\mathrm{lab}}=0$, we compare CPUTS with classical conformal prediction based solely on source $P$ samples (CP-P), implemented using the \textit{conformal.pred.split} function from the \textit{conformalInference} R package. Tables~\ref{tab:np_setting1} and~\ref{tab:np_setting2} report the results under the location shift and location--scale shift, respectively.
CP-P directly uses source data for calibration without accounting for the distribution shift, which may lead to inaccurate coverage for the target population. Under the location shift, CP-P exhibits substantial undercoverage. In contrast, under the location--scale shift, where the target response distribution is more concentrated than the source distribution, CP-P produces overly conservative prediction intervals.
By reweighting the source calibration samples, CPUTS accounts for the distribution shift and achieves coverage close to the nominal 90\% level across both settings. As $n_P$ increases, the average interval length decreases, while coverage remains close to the nominal level.

In the semi-supervised setting with $n_Q^{\mathrm{lab}}>0$, we compare CPUTS with two classical conformal prediction methods: CP-P using only source $P$ samples and CP-Q using only labeled target $Q$ samples. Tables~\ref{tab:nq_setting1} and~\ref{tab:nq_setting2} report the results under the location shift and location--scale shift, respectively. 
Similar to the unsupervised setting, CP-P exhibits undercoverage or overcoverage due to the distribution shift. Although CP-Q achieves coverage close to the nominal level, it tends to produce relatively long prediction intervals when labeled target samples are limited. By incorporating information from the source data, CPUTS maintains coverage close to the nominal level while improving predictive efficiency, particularly when $n_Q^{\mathrm{lab}}$ is small. For example, under the nonlinear model with a location shift and $n_Q^{\mathrm{lab}}=20$, CPUTS reduces the average interval length by approximately 48\% compared with CP-Q. As $n_Q^{\mathrm{lab}}$ increases, CP-Q produces shorter intervals, narrowing or even reversing the efficiency advantage of CPUTS.

\begin{table}[!htbp]
\caption{Coverage probability (CovP, \%) and average length (AL, with standard error, SE) of 90\% conformal prediction intervals across two models for varying source sample sizes $n_P$, with $n_Q = 1000$. Results are averaged over 300 test points and 1000 replications under the unsupervised target setting ($n_Q^{\text{lab}}=0$) with a location shift.}
\centering
\label{tab:np_setting1}
\begin{tabular}{ccccc}
\toprule
& \multicolumn{2}{c}{CPUTS} & \multicolumn{2}{c}{CP-P} \\
\cmidrule(lr){2-3}\cmidrule(lr){4-5}
 $n_P$      & CovP & AL (SE) & CovP & AL (SE) \\
\midrule
\multicolumn{5}{c}{\textit{Linear model}} \\
500   & \textbf{90.8} & 2.540 (0.010) & 81.2 & 2.330 (0.004) \\
1000  & \textbf{90.2} & 2.317 (0.006) & 80.9 & 2.318 (0.003) \\
2000  & \textbf{89.9} & 2.232 (0.003) & 80.9 & 2.311 (0.002) \\
3000  & \textbf{89.8} & 2.212 (0.003) & 80.6 & 2.309 (0.002) \\
4000  & \textbf{89.6} & 2.196 (0.002) & 80.6 & 2.309 (0.002) \\
\midrule
\multicolumn{5}{c}{\textit{Nonlinear model}} \\
500  & \textbf{91.2} & 3.226 (0.015) & 78.6 & 2.943 (0.006) \\
1000 & \textbf{90.6} & 2.954 (0.008) & 78.3 & 2.932 (0.004) \\
2000 & \textbf{90.3} & 2.867 (0.006) & 78.2 & 2.933 (0.003) \\
3000 & \textbf{90.3} & 2.859 (0.005) & 78.2 & 2.941 (0.003) \\
4000 & \textbf{90.0} & 2.850 (0.005) & 78.1 & 2.944 (0.002) \\
\bottomrule
\end{tabular}
\begin{tablenotes}[para]
\centering
{\footnotesize
CPUTS: Proposed method;
CP-P: Conformal prediction using only $P$-distribution samples.}
\end{tablenotes}
\end{table}

\begin{table}[!htbp]
\caption{Coverage probability (CovP, \%) and average length (AL, with standard error, SE) of 90\% conformal prediction intervals across two models for varying source sample sizes $n_P$, with $n_Q = 1000$. Results are averaged over 300 test points and 1000 replications under the unsupervised target setting ($n_Q^{\text{lab}}=0$) with a location--scale shift.}
\centering
\label{tab:np_setting2}
\begin{tabular}{ccccc}
\toprule
 & \multicolumn{2}{c}{CPUTS} & \multicolumn{2}{c}{CP-P} \\
\cmidrule(lr){2-3}\cmidrule(lr){4-5}
 $n_P$      & CovP & AL (SE) & CovP & AL (SE) \\
\midrule
\multicolumn{5}{c}{\textit{Linear model}} \\
500  & \textbf{91.2} & 1.663 (0.008) & 95.3 & 2.702 (0.005) \\
1000 & \textbf{90.5} & 1.549 (0.006) & 95.3 & 2.683 (0.003) \\
2000 & \textbf{90.5} & 1.494 (0.005) & 95.4 & 2.676 (0.002) \\
3000 & \textbf{90.6} & 1.481 (0.005) & 95.5 & 2.674 (0.002) \\
4000 & \textbf{90.5} & 1.462 (0.005) & 95.4 & 2.672 (0.002) \\
\midrule
\multicolumn{5}{c}{\textit{Nonlinear model}} \\
500  & \textbf{93.5} & 2.388 (0.027) & 98.8 & 4.365 (0.010) \\
1000 & \textbf{91.9} & 2.036 (0.017) & 99.2 & 4.379 (0.008) \\
2000 & \textbf{90.8} & 1.855 (0.015) & 99.3 & 4.402 (0.006) \\
3000 & \textbf{90.5} & 1.789 (0.013) & 99.4 & 4.438 (0.006) \\
4000 & \textbf{90.2} & 1.754 (0.013) & 99.5 & 4.456 (0.005) \\
\bottomrule
\end{tabular}
\begin{tablenotes}[para]
\centering
{\footnotesize
CPUTS: Proposed method;
CP-P: Conformal prediction using only $P$-distribution samples.}
\end{tablenotes}
\end{table}

\begin{table}[!htbp]
\caption{Coverage probability (CovP, \%) and average length (AL, with standard error, SE) of 90\% conformal prediction intervals across two models for varying target sample sizes $n_Q = 1000 + n_Q^{\text{lab}}$, with $n_P = 2000$. Results are averaged over 300 test points and 1000 replications under the semi-supervised setting ($n_Q^{\text{lab}} > 0$) with a location shift.}
\centering
\label{tab:nq_setting1}
\begin{tabular}{ccccccc}
\toprule
 & \multicolumn{2}{c}{CPUTS} & \multicolumn{2}{c}{CP-P} & \multicolumn{2}{c}{CP-Q} \\
\cmidrule(lr){2-3}\cmidrule(lr){4-5}\cmidrule(lr){6-7}
 $n^{\text{lab}}_Q$ & CovP & AL (SE) & CovP & AL (SE) & CovP & AL (SE) \\
\midrule
\multicolumn{7}{c}{\textit{Linear model}} \\
20  & \textbf{90.0} & \textbf{2.383} (0.006) & 80.9 & 2.309 (0.002) & 90.7 & 2.970 (0.030) \\
50  & \textbf{89.8} & \textbf{2.301} (0.005) & 80.7 & 2.309 (0.002) & 92.4 & 2.778 (0.019) \\
100 & \textbf{89.9} & \textbf{2.269} (0.004) & 80.7 & 2.310 (0.002) & 90.2 & 2.427 (0.010) \\
300 & \textbf{89.9} & \textbf{2.231} (0.004) & 80.8 & 2.313 (0.002) & 90.1 & 2.346 (0.006) \\
500 & \textbf{89.9} & \textbf{2.220} (0.004) & 80.8 & 2.312 (0.002) & 90.1 & 2.330 (0.004) \\
\midrule
\multicolumn{7}{c}{\textit{Nonlinear model}} \\
20  & \textbf{89.8} & \textbf{2.966} (0.009) & 78.4 & 2.936 (0.003) & 91.0 & 5.748 (0.230) \\
50  & \textbf{90.0} & \textbf{2.824} (0.006) & 78.4 & 2.936 (0.003) & 92.5 & 3.807 (0.068) \\
100 & \textbf{90.3} & \textbf{2.809} (0.006) & 78.4 & 2.936 (0.003) & 90.1 & 3.025 (0.014) \\
300 & \textbf{90.3} & \textbf{2.796} (0.006) & 78.5 & 2.934 (0.003) & 89.9 & 2.925 (0.007) \\
500 & \textbf{90.2} & \textbf{2.787} (0.005) & 78.2 & 2.934 (0.003) & 90.0 & 2.920 (0.006) \\
\bottomrule
\end{tabular}
\begin{tablenotes}[para]
\centering
{\footnotesize
CPUTS: Proposed method;
CP-Q: Conformal prediction using only $Q$-distribution samples; \\
CP-P: Conformal prediction using only $P$-distribution samples.}
\end{tablenotes}
\end{table}

\begin{table}[!htbp]
\caption{Coverage probability (CovP, \%) and average length (AL, with standard error, SE) of 90\% conformal prediction intervals across two models for varying target sample sizes $n_Q = 1000 + n_Q^{\text{lab}}$, with $n_P = 2000$. Results are averaged over 300 test points and 1000 replications under the semi-supervised setting ($n_Q^{\text{lab}} > 0$) with a location--scale shift.}
\centering
\label{tab:nq_setting2}
\begin{tabular}{ccccccc}
\toprule
 & \multicolumn{2}{c}{CPUTS} & \multicolumn{2}{c}{CP-P} & \multicolumn{2}{c}{CP-Q} \\
\cmidrule(lr){2-3}\cmidrule(lr){4-5}\cmidrule(lr){6-7}
 $n^{\text{lab}}_Q$ & CovP & AL (SE) & CovP & AL (SE) & CovP & AL (SE) \\
\midrule
\multicolumn{7}{c}{\textit{Linear model}} \\
20  & \textbf{90.2} & \textbf{1.542} (0.006) & 95.4 & 2.675 (0.003) & 90.8 & 1.879 (0.019) \\
50  & \textbf{90.2} & \textbf{1.515} (0.005) & 95.4 & 2.676 (0.003) & 92.4 & 1.745 (0.011) \\
100 & \textbf{90.2} & \textbf{1.496} (0.005) & 95.4 & 2.676 (0.003) & 90.3 & 1.544 (0.006) \\
300 & \textbf{90.2} & \textbf{1.484} (0.005) & 95.5 & 2.678 (0.003) & 90.2 & 1.497 (0.003) \\
500 & \textbf{90.2} & 1.486 (0.005) & 95.5 & 2.677 (0.002) & 90.0 & 1.483 (0.003) \\
\midrule
\multicolumn{7}{c}{\textit{Nonlinear model}} \\
20  & \textbf{90.6} & \textbf{1.857} (0.014) & 99.3 & 4.402 (0.006) & 91.0 & 2.356 (0.056) \\
50  & \textbf{90.6} & \textbf{1.823} (0.015) & 99.3 & 4.405 (0.006) & 92.3 & 1.915 (0.019) \\
100 & \textbf{90.6} & 1.797 (0.015) & 99.3 & 4.407 (0.006) & 90.2 & 1.632 (0.007) \\
300 & \textbf{90.8} & 1.795 (0.015) & 99.3 & 4.405 (0.006) & 90.0 & 1.575 (0.004) \\
500 & \textbf{90.6} & 1.777 (0.014) & 99.3 & 4.407 (0.006) & 90.0 & 1.568 (0.003) \\
\bottomrule
\end{tabular}
\begin{tablenotes}[para]
\centering
{\footnotesize
CPUTS: Proposed method;
CP-Q: Conformal prediction using only $Q$-distribution samples; \\
CP-P: Conformal prediction using only $P$-distribution samples.}
\end{tablenotes}
\end{table}

\begin{figure}[!htbp]
    \centering
    \includegraphics[width=\linewidth]{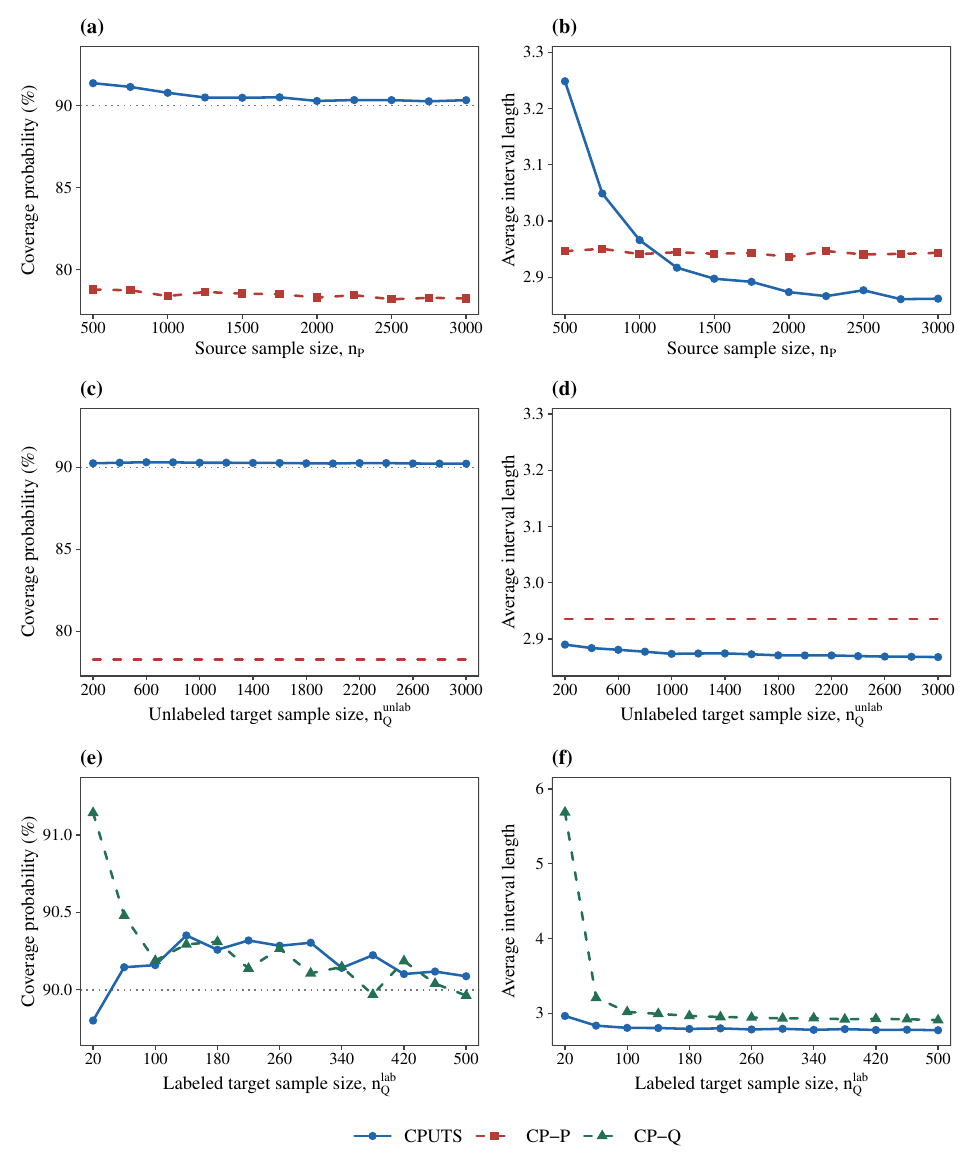}
    \captionsetup{
        width=0.9\linewidth,
        justification=justified,
        singlelinecheck=false
    }
    \caption{Coverage probability and average length
    of 90\% conformal prediction intervals under the nonlinear
    model with a location shift.
    Top row: unsupervised setting with varying $n_P$
    and fixed $n_Q^{\mathrm{unlab}}=1000$.
    Middle row: unsupervised setting with fixed $n_P=2000$
    and varying $n_Q^{\mathrm{unlab}}$.
    Bottom row: semi-supervised setting with fixed $n_P=2000$,
    $n_Q^{\mathrm{unlab}}=1000$, and varying
    $n_Q^{\mathrm{lab}}$.}
    \label{fig:plot_nPQ}
\end{figure}

In addition, we conduct additional simulations to examine how different sample sizes affect the performance of CPUTS. Using the nonlinear model with a location shift as an illustrative example, we present the results in Figure~\ref{fig:plot_nPQ}.
In the unsupervised setting, CPUTS maintains coverage close to the nominal level while its average interval length decreases as $n_P$ increases. This improvement may be attributed to more stable estimation of the density-ratio weights and nonconformity scores, together with reduced variability in weighted calibration. Meanwhile, CP-P continues to exhibit substantial undercoverage despite the increasing source sample size.
When varying the number of unlabeled target samples $n_Q^{\mathrm{unlab}}$ while fixing $n_P=2000$, CPUTS maintains nearly unchanged coverage probability and average interval length, suggesting stable performance over the range considered.
Finally, in the semi-supervised setting, CPUTS maintains coverage close to the nominal level and consistently produces shorter prediction intervals than CP-Q, particularly when labeled target samples are limited. This advantage reflects the use of additional source information, while the efficiency gap gradually narrows as $n_Q^{\mathrm{lab}}$ increases and CP-Q benefits from more labeled target data.

\section{Data Application}\label{sec:real_data}
To evaluate the real-world performance of our proposed method, we analyze the Medical Information Mart for Intensive Care III (MIMIC-III), a freely available database developed by the MIT Laboratory for Computational Physiology (\href{https://doi.org/10.13026/C2XW26}{https://doi.org/10.13026/C2XW26}), as described in \cite{johnson2016mimic}.
MIMIC-III contains health records for over 40,000 patients admitted to critical care units at Beth Israel Deaconess Medical Center between 2001 and 2012, including demographics, laboratory measurements, and vital signs.
To predict a patient’s condition during ICU stay and assess the extent of organ dysfunction or failure, we use the Sequential Organ Failure Assessment (SOFA) score as the response variable $Y$.
The SOFA score consists of six components representing the function of specific organ systems: respiratory, cardiovascular, hepatic, coagulation, renal, and neurological. The total score ranges from 0 (best) to 24 (worst).
Following \cite{lee2024doubly}, we select covariates based on two criteria: an absolute correlation with the outcome $Y$ greater than 0.2 and a missing rate less than 1\%.
As a result, we select 16 covariates, including six vital signs (such as blood pressure, respiratory rate, oxygen saturation, and body temperature) and ten laboratory measurements (such as glucose, urea nitrogen, platelets, red blood cell count, and hematocrit).
Following the data cleaning and preprocessing procedures in \cite{lee2024doubly}, we obtained $n = 14{,}080$ records.

To investigate potential target shift across insurance types, we define the source population $P$ as patients with private, government, or self-pay insurance ($n_P = 9{,}580$) and the target population $Q$ as those covered by Medicaid or Medicare ($n_Q = 4{,}500$).
A Kolmogorov–Smirnov test indicates a significant difference in the marginal distribution of $Y$ between $P$ and $Q$, with $p$-value $<10^{-15}$, suggesting that $p(y) \neq q(y)$.
To assess whether $p(\mathbf{x}\,|\, y) = q(\mathbf{x}\,|\, y)$, we use the invariant environment prediction test \citep{heinze2018invariant} from the R package \textit{CondIndTests} to test the conditional independence between each covariate and insurance type given $Y$. The resulting $p$-values, ranging from 0.12 to 0.79,
do not indicate significant differences in the conditional
distributions of individual covariates between the two populations.
We observe that the five racial groups, White ($n = 10{,}804$), Black ($n = 1{,}538$), Asian ($n = 410$), Hispanic ($n = 767$), and Other ($n = 561$), exhibit different relationships with the outcome $Y$. 
However, race cannot be used as a covariate because the target shift assumption $p(\text{race}\,|\, y)=q(\text{race}\,|\, y)$ is violated; a Cochran–Mantel–Haenszel test strongly rejects the conditional independence between race and insurance type given $Y$, with $p$-value $<10^{-15}$. 
We therefore perform predictions separately for each racial group. Within each racial group, the conditional-independence tests
for individual covariates do not detect evidence
against the target-shift assumption.

\begin{figure}[htpb!]
    \centering
    \includegraphics[width=0.8\linewidth]{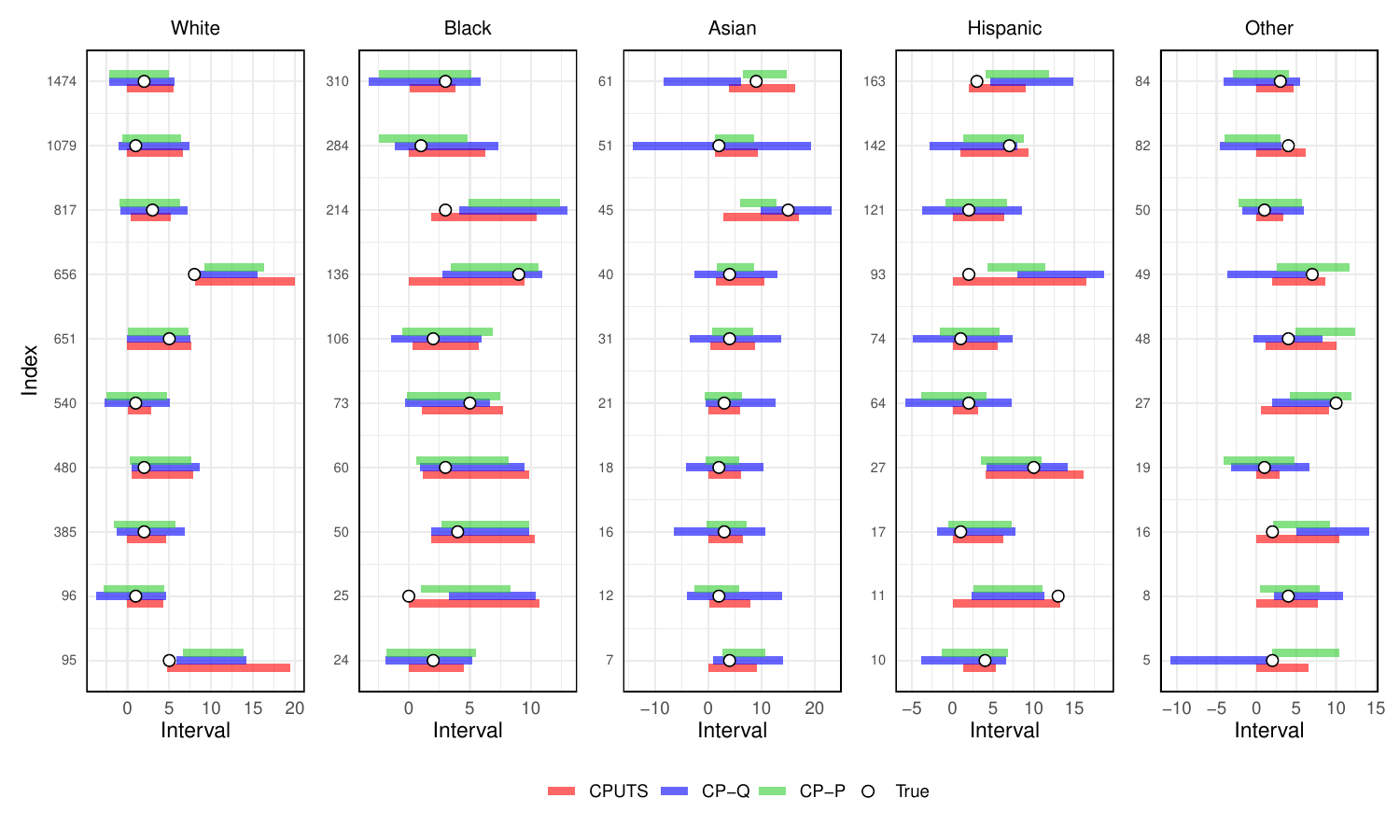}
    \caption{Comparison of 90\% prediction intervals from CPUTS, CP-Q, and CP-P for selected subjects from the five racial groups, with subject indices shown on the vertical axis.}
    \label{fig:each_subject}
\end{figure}

To mimic a semi-supervised setting, we mask a subset of target outcomes and use the remaining labeled target data to improve prediction efficiency.
We then apply our method CPUTS to predict SOFA scores for the target patients covered by Medicaid and Medicare in the semi-supervised target scenario, and compare it with two classical conformal prediction methods, CP-Q and CP-P, as described in Section~\ref{sec:simul}.
To stabilize quantile estimation for the ordinal outcome $Y$, we add a small random perturbation to $Y$ before computing the quantiles.
Following the parameter settings in Section~\ref{sec:simul}, we obtain the results summarized in Table~\ref{tab:coverage-length}, which report predictive performance across the different racial groups, averaged over all test points from leave-one-out cross-validation.
The empirical coverage probabilities of CPUTS and CP-Q
are generally close to the nominal level, whereas CP-P
exhibits undercoverage.
Moreover, CPUTS has a smaller average interval length
than CP-Q in each racial group, particularly when the
number of target samples is limited. For example, in the Asian subgroup, CPUTS produces shorter intervals for 94.4\% of test points compared with CP-Q, highlighting the effectiveness of the proposed method. We also present prediction intervals for selected patients from the five racial groups in Figure~\ref{fig:each_subject}, which support the same conclusion.

\begin{table}[htpb!]
\centering
\small
\renewcommand\arraystretch{0.8}
\caption{Coverage probability (CovP, \%) and interval length (AL, with standard error, SE) of 90\% conformal prediction intervals across racial groups, averaged over all leave-one-out samples.}
\label{tab:coverage-length}
\begin{tabular}{lccccc}
\toprule
Race & Method & Sample Size & CovP\% & AL (SE) & $|$CPUTS$|<|$CP-Q$|$\\ 
\midrule
White   & CPUTS & 10804    & \textbf{90.1} & \textbf{7.349}(0.060) & 68.0\% \\
        & CP-Q  & 3098  & 89.9 & 8.081(0.007) & -- \\
        & CP-P  & 7706  & 87.0 & 7.174(0.003) & -- \\
\midrule
Black   & CPUTS & 1538    & \textbf{89.2} & \textbf{7.383}(0.124) & 70.7\% \\
        & CP-Q  & 738   & 92.1 & 8.176(0.029) & -- \\
        & CP-P  & 800   & 87.8 & 7.464(0.020) & -- \\
\midrule
Asian   & CPUTS & 410    & \textbf{91.5} & \textbf{7.595}(0.276) & 94.4\%\\
        & CP-Q  & 141   & 91.5 & 13.286(0.436) & -- \\
        & CP-P  & 269   & 88.7 & 7.503(0.081) & -- \\
\midrule
Hispanic& CPUTS & 767    & \textbf{89.3} & \textbf{7.906}(0.188) & 84.8\% \\
        & CP-Q  & 356   & 89.3 & 10.399(0.109) & -- \\
        & CP-P  & 411   & 85.4 & 7.734(0.039) & -- \\
\midrule
Other   & CPUTS & 561    & \textbf{90.5} & \textbf{6.952}(0.220) & 79.8\%\\
        & CP-Q  & 167   & 85.7 & 9.427(0.187) & -- \\
        & CP-P  & 394   & 83.3 & 7.986(0.065) & -- \\
\bottomrule
\end{tabular}
\begin{tablenotes}[para]
\centering
{\footnotesize 
CPUTS: Proposed method; 
CP-Q: Conformal prediction using only $Q$-distribution samples; \\
CP-P: Conformal prediction using only $P$-distribution samples; 
$|$CPUTS$| < |$CP-Q$|$: Proportion of instances in which the prediction interval length from CPUTS is shorter than that from CP-Q.}
  \end{tablenotes}
\end{table}

\section{Conclusion}
\label{sec:conc}
In this work, we propose a new CPUTS framework to address the target shift problem between source and target distributions, covering two cases: when the target distribution has no labels or only a few labeled samples.
Under the stated assumptions and tuning conditions, the method
achieves asymptotic marginal coverage of at least $1-\alpha$,
as established by Theorems~\ref{th:cover} and~\ref{thm:weight_rate},
while leaving room for further improvement.
First, its prediction interval efficiency could be improved by incorporating feature-based kernel weighting into the $p$-value in \eqref{p-value-task1}, enabling subgroup-level or conditional coverage guarantees.
Second, although the coverage depends on the accuracy of the weight estimation, and this error can be controlled, stronger properties may be achievable, such as the doubly robust property, under which accurate estimation of either the weights or the conditional density alone would suffice.
In addition, our framework uses a single source dataset to improve prediction for the target data. In practice, however, multiple source datasets with different distributions are often available, and integrating them to enhance predictive efficiency is an interesting direction for future work.

\section*{Supplementary Materials}
The online Supplementary Materials provide detailed technical proofs and sensitivity analyses.

\section*{Acknowledgment}
Tang's research has been partially supported by the National Natural Science Foundation of China grant 12371265, Natural Science Foundation of Shanghai grant 24ZR1455200, and Wang's research by the National Science Foundation grant DMS-2436216. 
During the preparation of this work, the authors used
ChatGPT (OpenAI) to assist with manuscript checking and language polishing. The authors reviewed and edited the AI-assisted content as necessary and take full responsibility for the content of the publication.

\bibliographystyle{biometrika}
\bibliography{references}

\end{document}